\documentclass[authoryear, 12pt, 3p, letter]{elsarticle} 
\usepackage{amssymb}
\usepackage{amsmath}
\usepackage{rotating}

\usepackage{ textcomp }
\usepackage{comment}
\usepackage{enumerate}
\usepackage{subfigure}
\usepackage{lipsum}
\usepackage{color}
\usepackage{multirow}
\usepackage{lineno}

\journal{Icarus}

\begin{document}

\begin{frontmatter}

\title{Possible ground fog detection from SLI imagery of Titan}

\author[addy1]{Christina L Smith}
\ead{chrsmith@yorku.ca}
\author[addy1]{Brittney A. Cooper}
\author[addy1]{John E. Moores}
\address[addy1]{Centre for Research in Earth and Space Science, York University, 4700 Keele Street, Toronto, M3J1P3, ON, Canada}\label{addy1}

\begin{abstract}

Titan, with its thick, nitrogen-dominated atmosphere, has been seen from satellite and terrestrial observations to harbour methane clouds. To investigate whether atmospheric features such as clouds could also be visible from the surface of Titan, data taken with the Side Looking Imager (SLI) on-board the Huygens probe after landing have been analysed to identify any potential atmospheric features. In total, 82 SLI images were calibrated, processed and examined for features. The calibrated images show a smooth vertical radiance gradient across the images, with no other discernible features. After mean-frame subtraction, six images contained an extended, horizontal feature that had a radiance value that lay outside the 95\% confidence limit of the predicted radiance when compared to regions higher and lower in the images. The change in optical depth of these features were found to be between 0.005 and 0.014. It is considered that these features most likely originate from the presence of a fog bank close to the horizon that rises and falls during the period of observation.

\end{abstract}


\end{frontmatter}

\section{Introduction}

Titan is a unique body in the Solar System as the only moon harbouring a thick atmosphere (see discussions and further details in \citealp{Roe2012} and \citealp{Lorenz2008} and references therein). This atmosphere is composed primarily of nitrogen ($>90\%$), followed by methane (5\%) and smaller fractions of other components \citep{Roe2012, Gibbard1999}. The atmospheric pressure at the surface was measured by the Huygens Atmosphere Structure Instrument (HASI) on-board the Huygens Probe to be $1.467(1)\times10^5$ Pa \citep{Fulchignoni2005}. The lack of oxygen in the atmosphere of Titan allows for the production of complex organic molecules, such as poly aromatic hydrocarbons (PAHs) \citep{Waite2007}.  Complex hydrocarbon aerosols, commonly known as tholins, are thought to result in the dense orange haze that is characteristic of Titan's atmosphere and obscures the surface from view at visible wavelengths. The optical depth of the atmospheric haze is dependent on both wavelength and altitude: optical depth varies for a given point between 2 and 4.5 between $\sim450-950$ nm and can increase by a factor of 2 between 150 km and the surface \citep{Tomasko2005}.

Methane cycling, similar to the hydrological-cycling occurring on Earth, has been hypothesised to occur on Titan. \citet{Tokano2001} carried out in-depth 3-dimensional global circulation modelling of the tropospheric methane cycle, showing that methane cycling predominantly occurs within Titan's atmosphere. Surface darkening on Titan, consistent with surface wettening by methane precipitation, has been observed (\citealp{Barnes2013,Turtle2011}). Methane clouds have been detected from both terrestrial telescope observations (e.g. \citealp{Griffith2000, Roe2005}) and satellite observations (e.g. \citealp{Brown2010, Rodriguez2011}) at a variety of altitudes. Transient methane condensation clouds were also detected by \citet{Griffith1998}, at altitudes of 15 km. Cloud features have been detected from the surface of other planetary bodies (e.g. \citealp{Moores2015}), thus it is possible that cloud features could also be observed from the surface of Titan.

The Huygens probe, carried on-board the Cassini Spacecraft, arrived at Saturn on July 1st 2004 after a seven year journey from Earth. The Huygens probe was released from Cassini on December 25th 2004 and began its descent through the atmosphere of Titan on January 14th 2005. The 
probe descended through the atmosphere for 2.5 hours before landing on Titan's surface. The Huygens probe continued to take measurements and relay data for a further hour after landing \citep{Tomasko2005}.

In this paper, we present the results of an investigation of data taken with the Descent Imager/Spectrometer Radiometer (DISR, \citealp{Tomasko2002}) on-board the Huygens probe to search for features, such as clouds and mirages, in the lower atmosphere of Titan as viewed from the surface. 

\section{Materials and Methods}

\subsection{Instrument and data}

The DISR on-board Huygens is made up of a number of different sub-instruments to collect a variety of measurements at optical wavelengths. Three of these sub-instruments are imagers: the High Resolution Imager (HRI), the Medium Resolution Imager (MRI) and the Side Looking Imager (SLI). These three imagers observe in different directions and have differing resolutions and fields of view: the SLI has the largest field of view with the lowest resolution and the HRI has the smallest field of view with the highest resolution. A summary of the specifications of these imagers can be found in Table 2 of \citet{Tomasko2002}. 

The three aforementioned imagers share a $512\times520$ pixel silicon CCD detector with three other sub-instruments (UVLS, DVLS, SA Radiometer). Half of the CCD is masked and used as a ``memory region". Of the remaining half, the majority is used by the three imagers. A schematic layout of the CCD detector can be found in Fig. 1 of \citet{Tomasko2002}. 

Throughout the descent, the imagers took observations and continued to take images once the probe had landed on the surface. The total number of images received from the probe from the HRI, MRI and SLI instruments was 601. 

Only images taken with the Side Looking Imager (SLI) were analysed, as data taken with the High Resolution Imager (HRI) and Medium Resolution Imager (MRI) did not include images of the sky. As detailed in \citet{Tomasko2002}, the SLI had a Nadir range of 45.2-96.0\textdegree \,with an azimuth angle range of 25.6\textdegree. The field of view includes 6\textdegree \, of image above the horizon. The spectral range of all three imagers was 660-1000nm and the SLI had a pixel format of $128\times256$ pixels with a spatial scale of 0.2\textdegree \, per pixel.

The data from all three imagers were retrieved from the European Space Agency (ESA) Planetary Science Archive\footnote{ftp://psa.esac.esa.int/pub/mirror/CASSINI-HUYGENS/DISR/HP-SSA-DISR-2-3-EDR-RDR-V1.0/DATA/IMAGE/TABLE\_FORMAT/}. The SLI data were separated using the accompanying header files and those taken after landing - totalling 82 images - extracted manually.

The data retrieved from the Planetary Science Archive are available in two forms, binary tables ($128\times256$ pixels) and binary image files ($256\times512$ pixels). For the remainder of the paper, the data referred to are those from the binary table files. 

\subsection{Calibration}

The SLI data underwent a number of calibration and processing steps both on-board Huygens and on the ground prior to the data archiving. These steps are described in detail in the calibration report provided with the archive data \citep{CalibrationReport} and will be summarised here.

On board Huygens, a bad pixel map was used to replace the value of bad pixels with the value of another pixel in the same row. A flat-field correction was applied to remove a number of image artifacts, including a ``chicken wire" pattern caused by the optical fibre unit. The data were taken in 12 bits per pixel format, but were only able to be transmitted at 8 bits per pixel. Thus, square root processing was used to map the 12 bit data to 8 bit data. For the MRI and HRI instruments, the algorithm was implemented in an adaptive form to improve the representation of the data, but as the SLI images included the sky, this was not applied. The image data contained hot pixels which would have had a detrimental impact on the quality of the transmitted data, therefore hot pixels were replaced. Prior to transmission, the data were compressed in $16\times16$ pixel blocks using a discrete cosine function.

After the data were received on Earth, further processing took place. The 8 bits per pixel data were mapped back to 12 bits per pixel using a reverse square root processing algorithm. The images were also decompressed, although the compression carried out on-board is not completely reversible.

The data were retrieved from the Planetary Science Archive in this state. In order to retrieve radiometrically calibrated data for use in further analyses, further calibration processes were required - these are also explained in further detail in \citet{CalibrationReport}. The dark current after the initial 40 minutes of descent is negligible \citep{Tomasko2002}, therefore a value of 8 data numbers (DN) was added to each pixel to undo the estimated dark current removal on-board. The exposure times of the observations varied between 2 and 50 ms; to compensate for this, each of the images was divided by its respective exposure time. The absolute responsivity, $A$, at the CCD instrument temperature (ranging between approximately 170-185 K) at the time each dataset was collected was computed using the model found in \citet{CalibrationReport}:

\begin{equation}
\begin{split}
A= -3729544.2+78230.288T-501.89048T^2\\+1.4119032T^3-0.0014748198T^4
\end{split}
\end{equation}

where $T$ is the CCD temperature at the time of observation (K) and $A$ is in units of counts s$^{-1}$ W$^{-1}$ m$^2$ Sr.

After processing and calibration is complete, the pixel values are returned in units of radiance within the observing band of the detector.

\subsection{Analysis}\label{procs}

The final image set comprises of 82 calibrated images; these were analysed in a number of ways to identify the presence of atmospheric features. Animated gifs were created at each point to aid in manual feature identification. All combinations of the following procedures were applied to the data in order to identify the presence, significance and potential origin of atmospheric features. 

\begin{itemize}

\item The images were cropped to the region between 4 and 40 pixels from the upper edge and 7 and 14 pixels from the left and right edges respectively. This region constitutes the portion of the original images containing the sky with edge-related artifacts removed.

\item A mean frame was produced from all images and subsequently subtracted from each image. This enhances edges and compression-related artifacts.

\item A Gaussian filter ({\sc{Python}}'s scipy Gaussian\_filter function with a kernel size of one to five) was applied to smooth the variations caused by compression artifacts.

\item Temporally consecutive images were co-added  in pairs  to reduce the image noise.

\item Difference imaging was applied to temporally consecutive pairs of images. The second image was subtracted from the first to identify any changes that had occurred between images.

\end{itemize}

\begin{figure*}
\centering

\addtolength{\subfigcapskip}{-5mm}

\subfigure[Cropped, calibrated image.]{\includegraphics[trim=1cm 4.0cm 0cm 5cm, clip=true, width=0.99\textwidth]{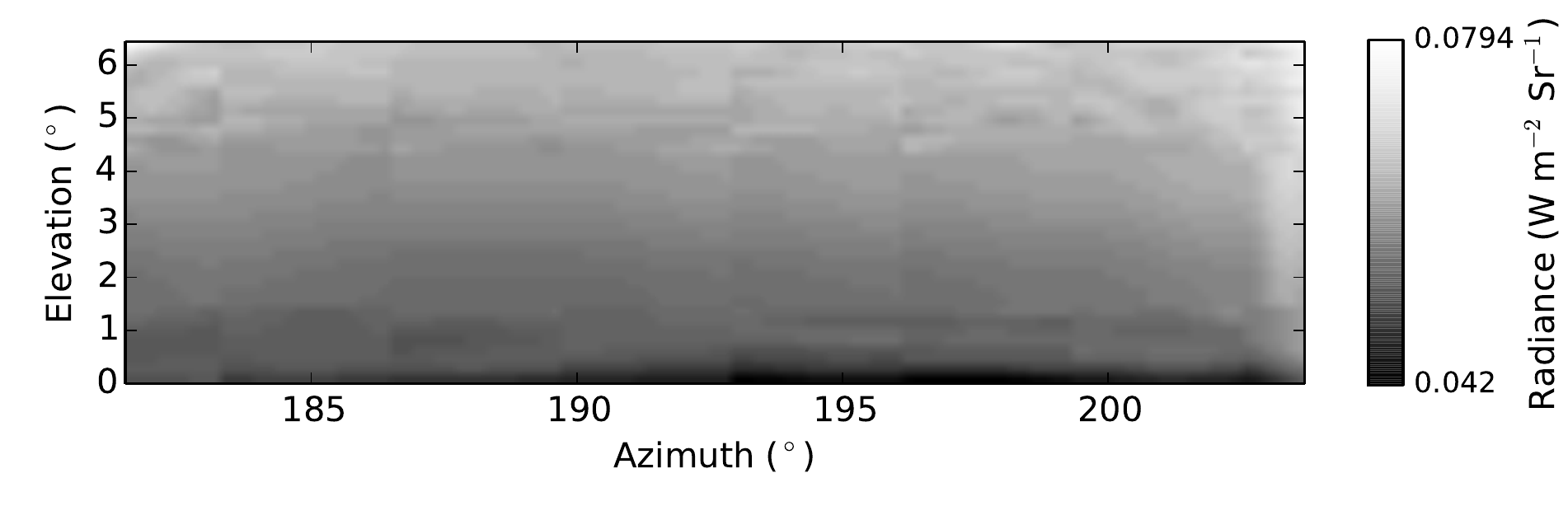}}

\subfigure[Cropped, calibrated image with blurring. Kernel size = 3.]{\includegraphics[trim=1cm 4.0cm 0cm 5cm, clip=true, width=0.99\textwidth]{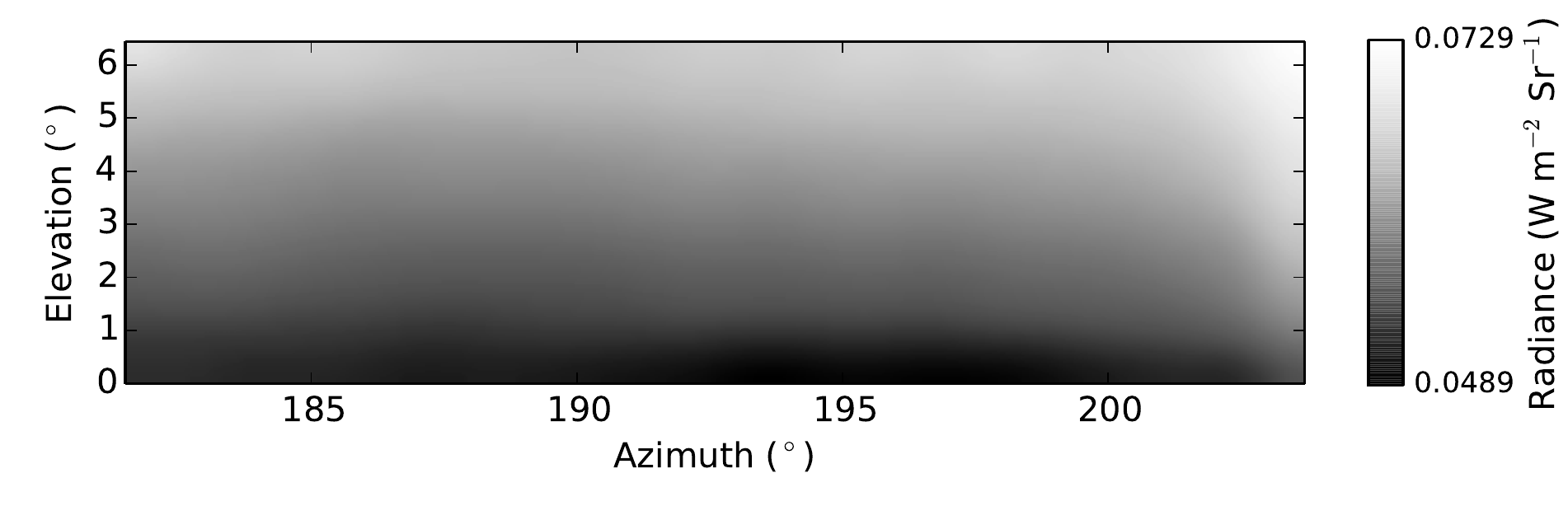}}

\subfigure[Cropped, calibrated, mean frame subtracted image.]{\includegraphics[trim=1cm 4.0cm 0cm 5cm, clip=true, width=0.99\textwidth]{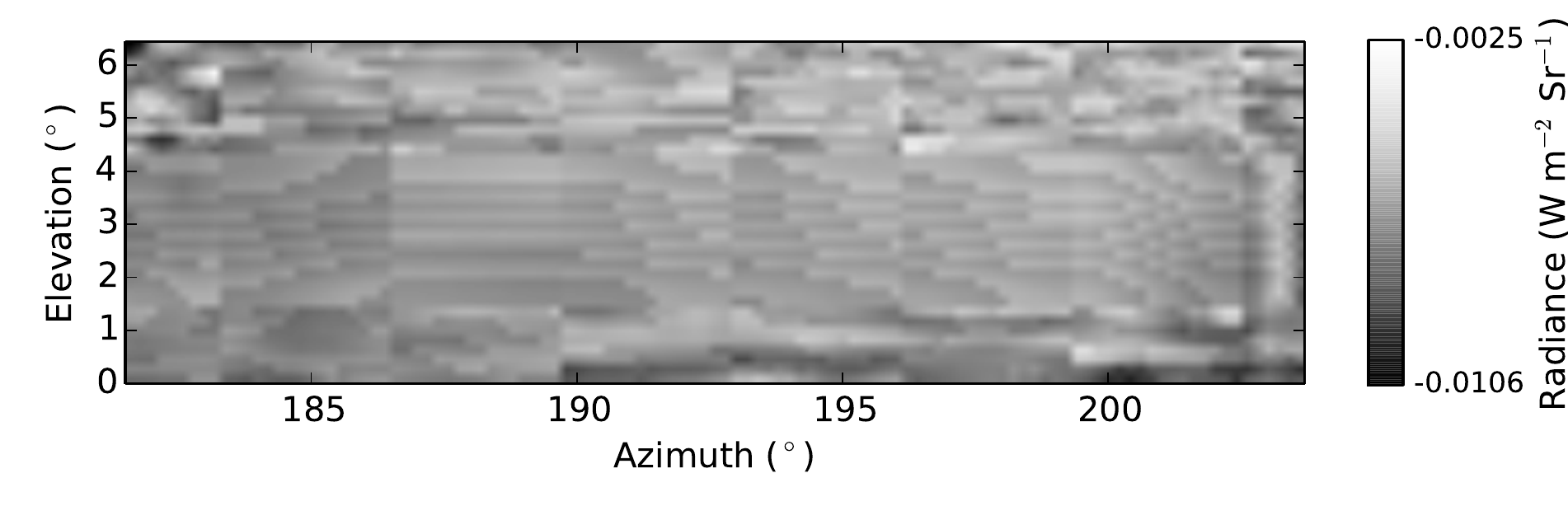}\label{mfsimage}}

\subfigure[Cropped, calibrated, mean frame subtracted image with blurring. Kernel size = 3.]{\includegraphics[trim=1cm 4.0cm 0cm 5cm, clip=true, width=0.99\textwidth]{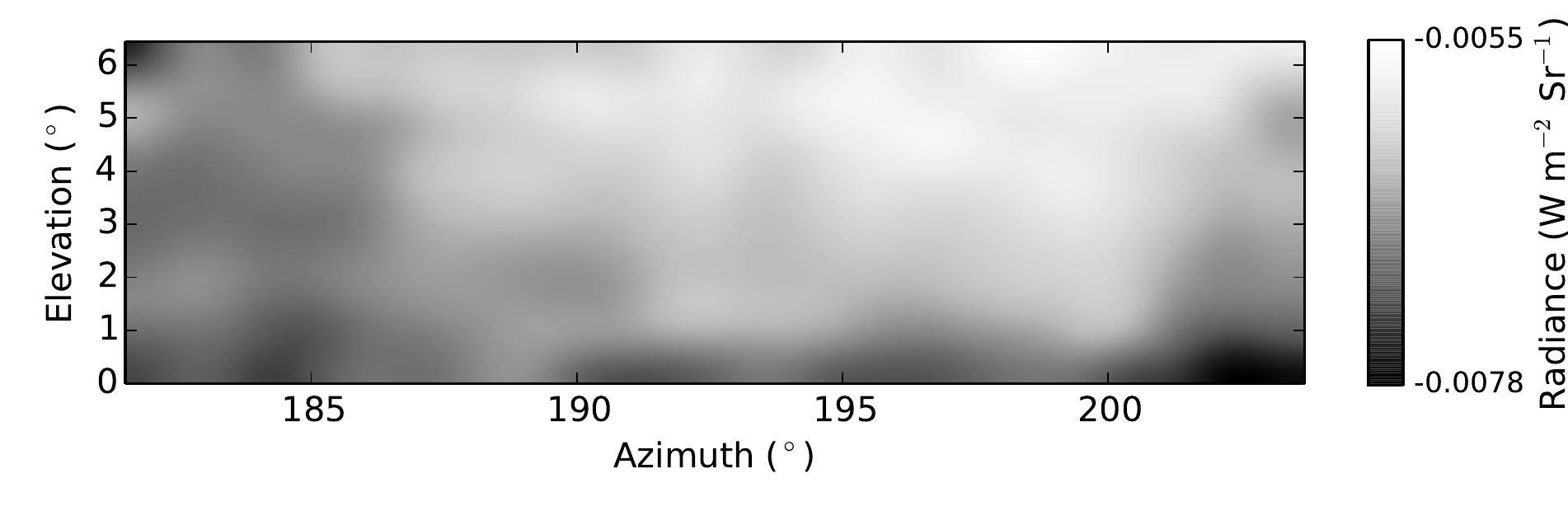}\label{sample}}

\caption{Cropped images with a selection of different processes applied, as described in Sect. \ref{procs}. All images are image number 0817, Mission Time 9755.38 seconds.}\label{diff_processes}
\end{figure*}

An example image with a selection of the aforementioned processes applied is shown in Fig. \ref{diff_processes}. The elevation angle has been measured assuming that the lower edge of the cropped image represents the horizon. The azimuth angle has been calculated under the assumption that the centre of the image is at an azimuthal angle of $193^{\circ}$ east of north \citep{Karkoschka2007}.  Unless otherwise stated, all analysis from this point onward is carried out on images which have been calibrated, cropped and (where mentioned) mean-frame subtracted and smoothed. Coadding and difference imaging gave no further detections on the data than those found using mean-frame subtraction from individual images and thus were not used further.

\section{Results}

\subsection{Radiance measurements}\label{rad_measurements}

The calibrated frames with no further processing show no distinct structure or evidence of clouds, mirages or artifacts. There is a general change in radiance vertically across all images. This variation is expected and has been observed on other planetary bodies  with atmospheres, such as Mars. 

\begin{figure*}
\centering
\addtolength{\subfigcapskip}{-3mm}
\setcounter{subfigure}{0}
\subfigure[Model fit from Equ. \ref{atm_equ}.]{\includegraphics[trim=0cm 0cm 3cm 15cm, clip=True,width=0.7\textwidth]{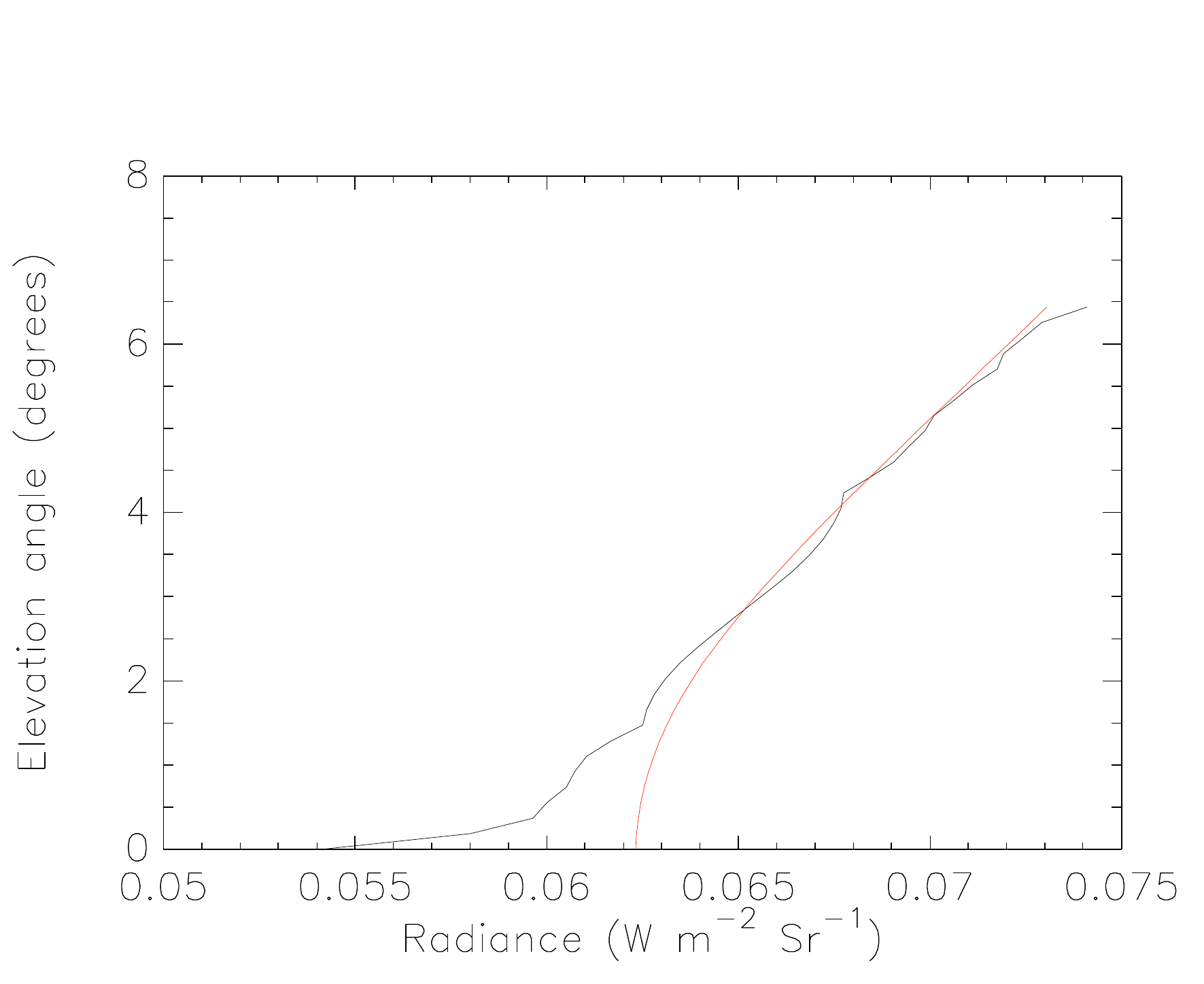}\label{beersfit_average}}

\subfigure[Calibrated mean-frame image.]{\includegraphics[trim=1cm 4.0cm 0cm 5cm, clip=true, width=0.99\textwidth]{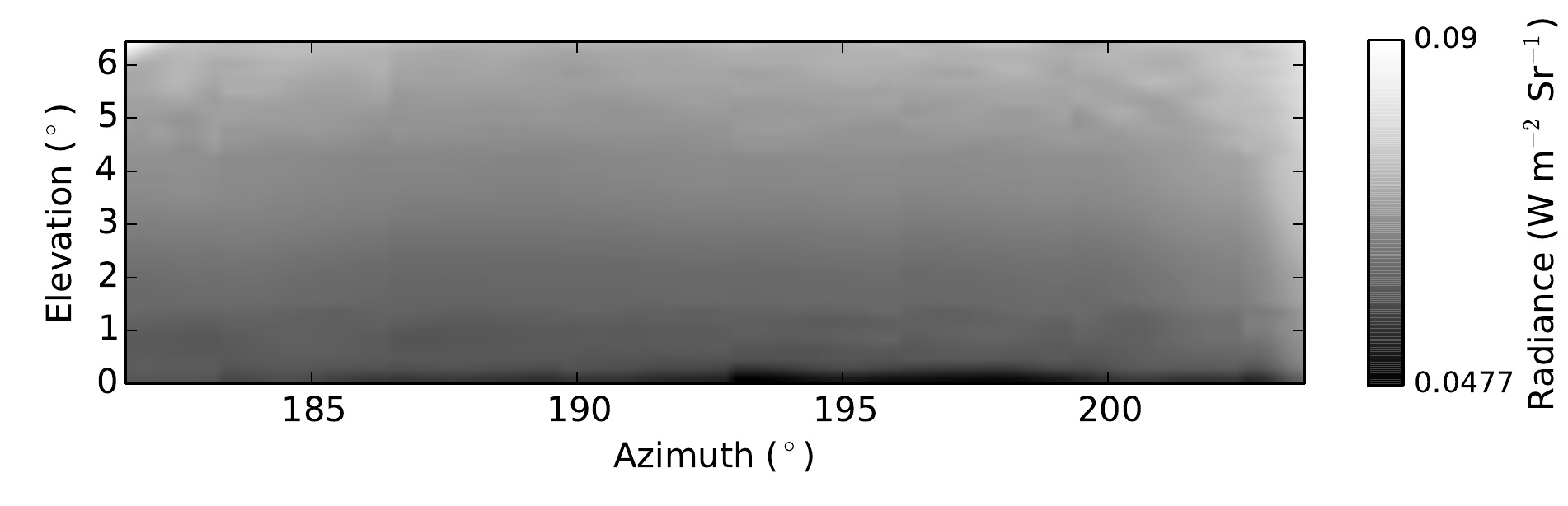}}
\caption{The lower figure shows the mean frame image with no smoothing. The upper image show the average vertical radiance profile (black line) along with a best-fit model using Equ. \ref{atm_equ} (upper plot).}\label{gradient_fits}
\end{figure*}

On Titan, the radiance increases with elevation angle making the sky darkest at the horizon. This is demonstrated in Fig \ref{gradient_fits}. The apparent horizon after landing was found to be $1-2^\circ$ above the theoretical horizon, taking into account the pitch of Huygens after landing - believed to be evidence of a small rise \citep{Karkoschka2007}. Assuming the multiple scattering in the atmosphere results in an absorption profile, this can be fitted using:

\begin{equation}\label{atm_equ}
I=\frac{F}{4\pi}\exp\left(-\frac{\tau}{\sin(\alpha+2^{\circ})}\right)+C
\end{equation}

\noindent where $F$ is the Solar flux (W m$^{-2}$) at the surface of Titan, $I$ is the radiance (W m$^{-2}$ Sr$^{-1}$), $\tau$ is the total vertical optical depth of the atmosphere, and $C$ is a constant. $\alpha$ is the measured elevation angle from the lower edge of the image and the addition of $2^{\circ}$ compensates for the apparent-to-true horizon discrepancy. 

The genetic algorithm {\sc{Pikaia}} \citep{Charbonneau1995} was used to fit the data and resulted in a $\tau=0.26316$, $F=0.81229$ and $C=0.06229$. The value of the Solar flux at the surface of Titan agrees well with the value of the Solar constant scaled to the distance of Titan at the time of the Huygens landing (9.053052 AU, as generated by the New Horizons Ephermeris\footnote{http://ssd.jpl.nasa.gov/horizons.cgi} at noon on January 14 2005) and attenuated to the surface with an optical depth of 3: $F_\text{scaled}=0.8292$ W m$^{-2}$. The model fit is shown in Fig. \ref{beersfit_average}. The model has good agreement with the data from an elevation of $2^{\circ}$ upwards. Below $2^{\circ}$, the model has a higher radiance than that of the data, suggesting that an additional absorption is occurring in this region, perhaps due to the presence of ground fog. Changing the horizon offset from $2^{\circ}$ shifts the model line vertically. Varying $C$ allows each horizon offset to be compensated for, allowing an acceptable fit to the data to be made whilst keeping all other parameters constant. However, as the reported apparent-to-true horizon discrepancy is $2^{\circ}$, this is the value used in this analysis.

\begin{figure*}
\centering
\addtolength{\subfigcapskip}{-3mm}
\subfigure[Mean frame subtracted image]{\includegraphics[trim=1cm 4.0cm 0cm 5cm, clip=true, width=0.99\textwidth]{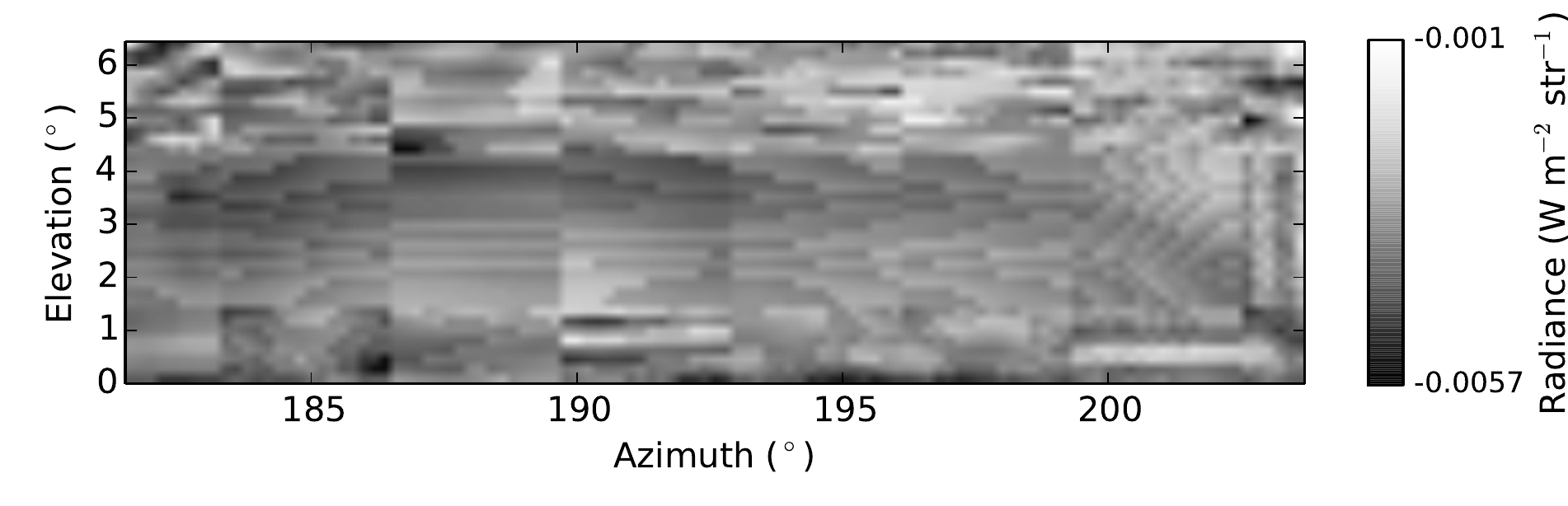}\label{mean_frame_subtracted}}
\subfigure[Mean frame subtracted image after the application of a Gaussian blur with kernel size 3.]{\includegraphics[trim=1cm 4.0cm 0cm 5cm, clip=true, width=0.99\textwidth]{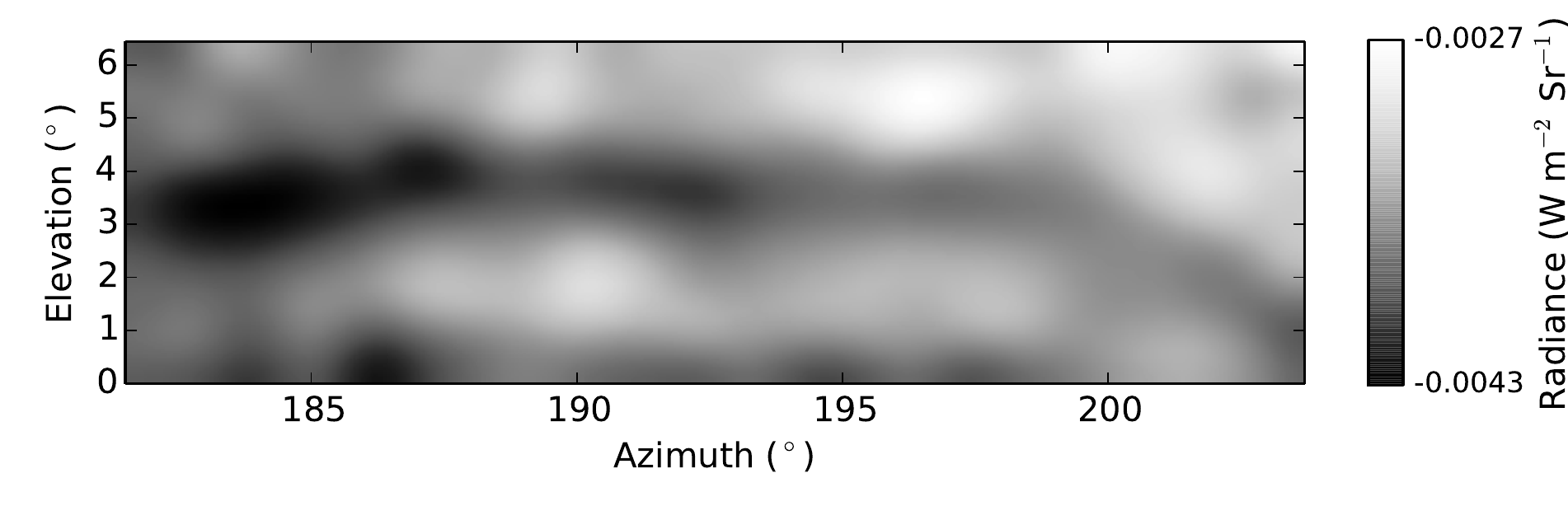}\label{mfs_gaussianblur}}
\caption{An example mean-frame subtracted image with a feature before and after Gaussian blurring.}
\end{figure*}

The large-scale gradient common across all images was removed using meanframe subtraction. This allowed small scale variations to be examined. However, as well as emphasising small image variations, mean frame subtraction has the side-effect of emphasising low-level image artifacts, which is shown clearly in Fig. \ref{mean_frame_subtracted}. These variations were suppressed using the Gaussian filter function in {\sc{python}} and the effect shown in Fig. \ref{mfs_gaussianblur}.

\begin{sidewaysfigure*}
\centering
\addtolength{\subfigcapskip}{-5mm}
\subfigure[Frame 0967, MT 11009.76 s.]{\includegraphics[trim=1cm 4.0cm 0cm 5cm, clip=true, width=0.49\textwidth]{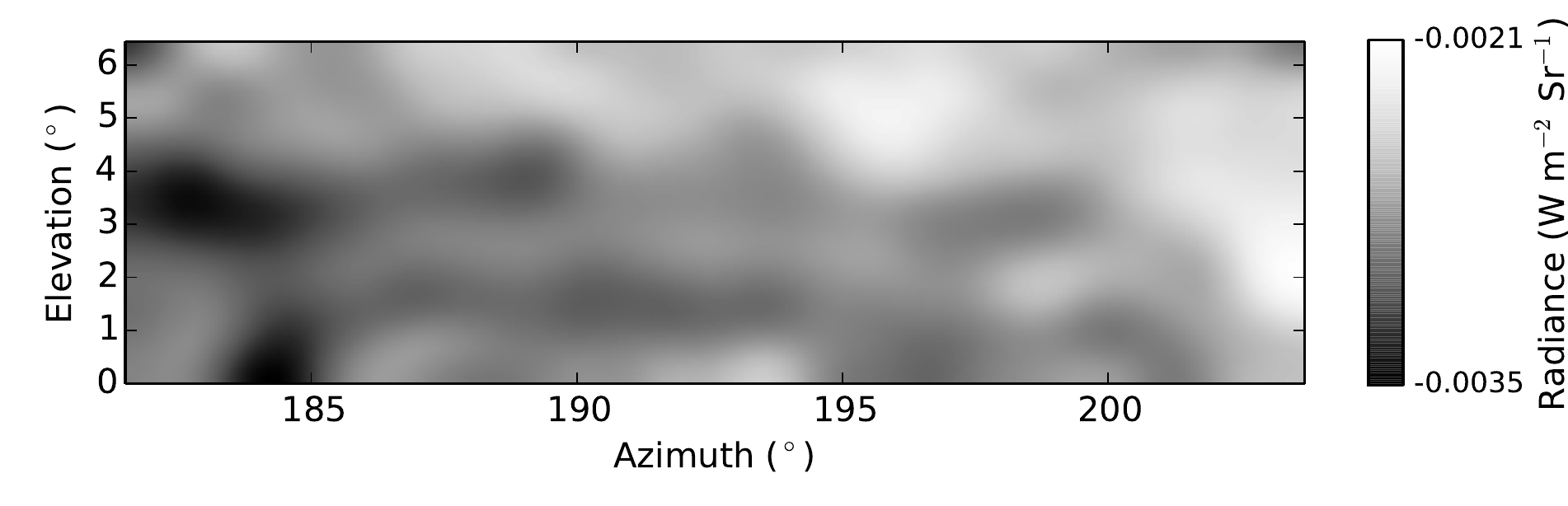}}
\setcounter{subfigure}{3}
\subfigure[Frame 1132, MT 12334.40 s.]{\includegraphics[trim=1cm 4.0cm 0cm 5cm, clip=true, width=0.49\textwidth]{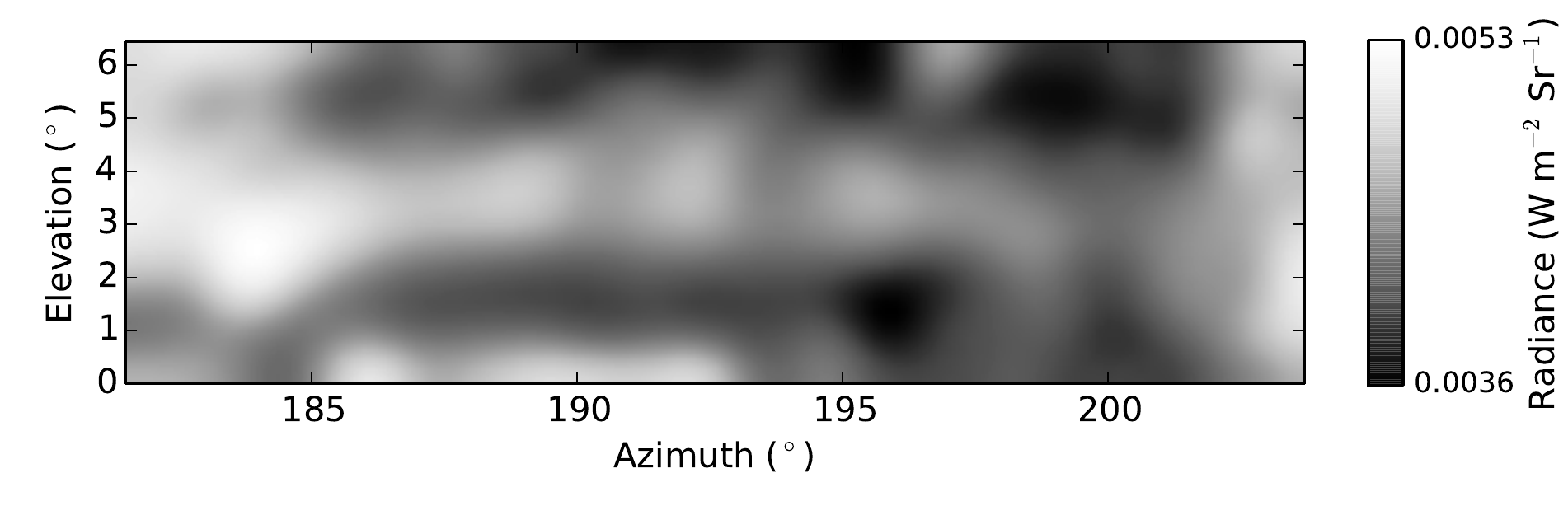}}
\medskip

\setcounter{subfigure}{1}
\subfigure[Frame 0970, MT 11013.18 s.]{\includegraphics[trim=1cm 4.0cm 0cm 5cm, clip=true, width=0.49\textwidth]{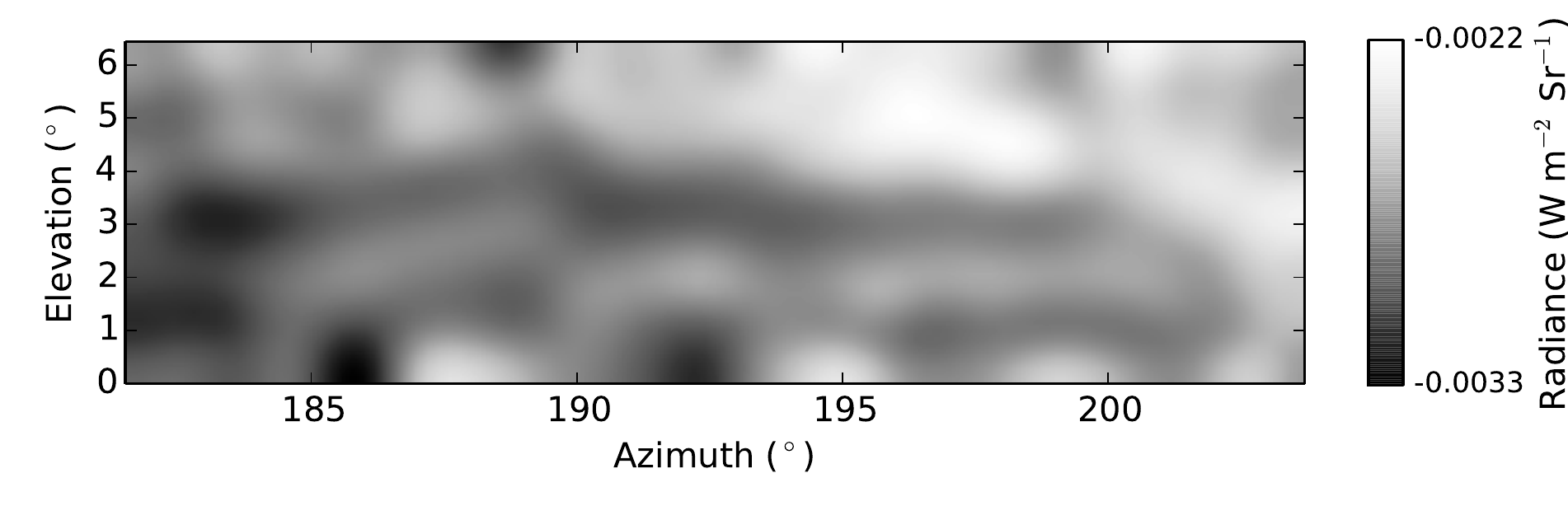}}
\setcounter{subfigure}{4}
\subfigure[Frame 1138, MT 12371.71 s.]{\includegraphics[trim=1cm 4.0cm 0cm 5cm, clip=true, width=0.49\textwidth]{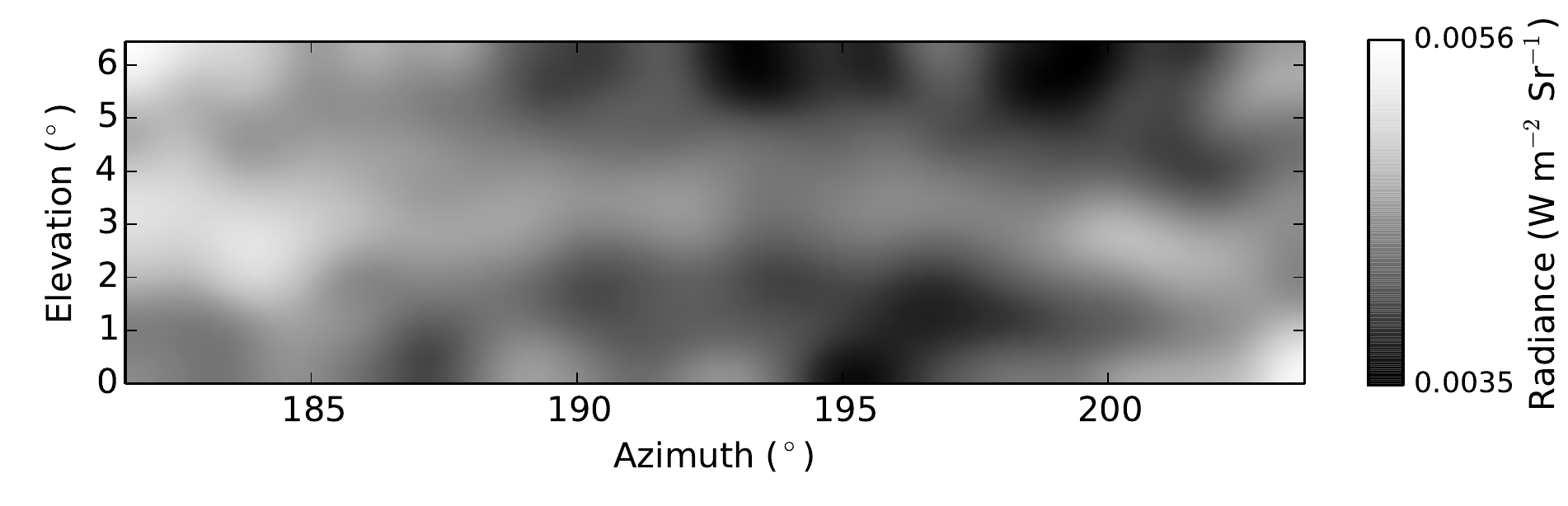}}
\medskip

\setcounter{subfigure}{2}
\subfigure[Frame 0979, MT 11185.34 s.]{\includegraphics[trim=1cm 4.0cm 0cm 5cm, clip=true, width=0.49\textwidth]{figures_for_paper/0979_forpaper_avsub_blurred.pdf}}
\setcounter{subfigure}{5}
\subfigure[Frame 1147, MT 12402.39 s.]{\includegraphics[trim=1cm 4.0cm 0cm 5cm, clip=true, width=0.49\textwidth]{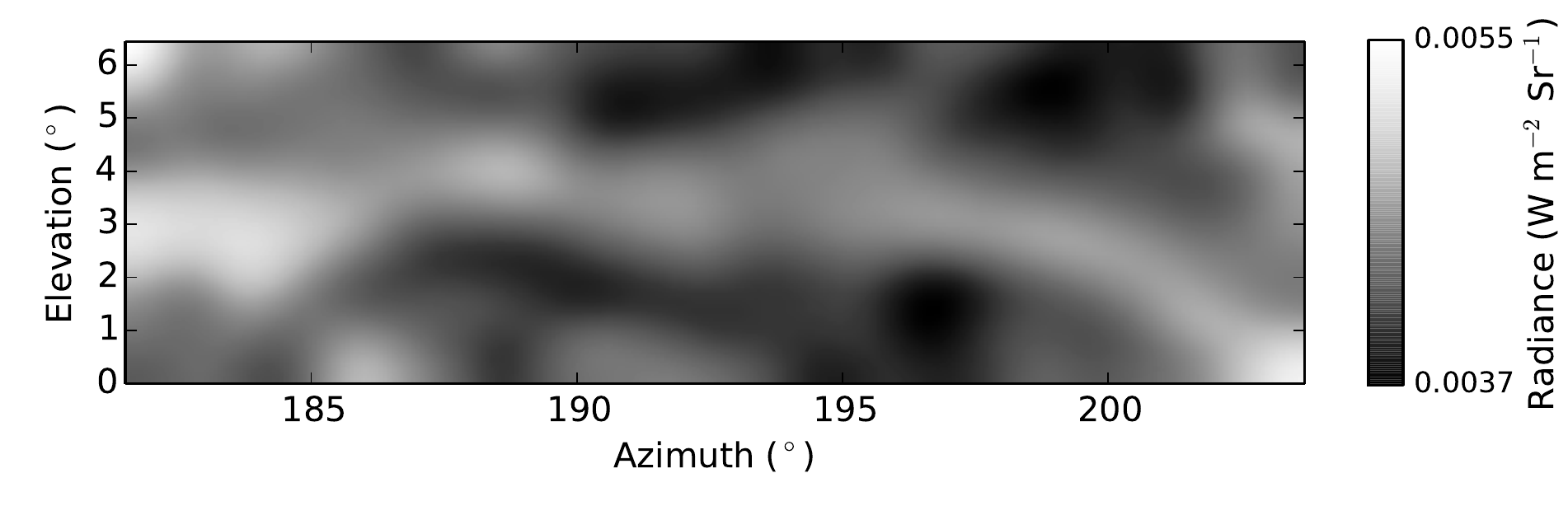}}
\medskip

\caption{All frames showing feature detections. All frames have been mean-frame subtracted and Gaussian blurring with a kernel size of 3 applied. The feature is seen as a dark horizontal band in the images (a) to (c) and as a light horizontal band in images (d) to (f).}\label{dets} 
\end{sidewaysfigure*}

Extended, horizontal linear features were detected in six images. These features were only detected after mean-frame subtraction, coaddition of frames or difference imaging was applied. Each frame with a detected atmospheric feature is shown in Fig. \ref{dets} after mean-frame subtraction and Gaussian blurring with a kernel size of 3. A representative frame with no detected feature is shown in Fig. \ref{sample}. The feature is seen as a horizontal band, either lighter or darker than the rest of the frame at approximately $3^{\circ}$ elevation.

The radiance of the feature was measured as the mean value in a $6\times6$ pixel square region, the centre of which was selected manually. The radiance of the background was taken as the average value in two $6\times6$ pixel square regions, the centres of which were vertically offset from the central pixel of the feature region by 10 pixels. The radiance measurements were carried out at three different azimuth angles across the image, shown for an example feature-detection image in Fig. \ref{optcalcreg}. These regions were selected as representative of the feature's high, low and medium radiance values. The difference in radiance between the feature region and the mean of the background regions is shown in Table \ref{RadTab}. The value in parentheses indicates the 95\% confidence limit on the final digit. This has been calculated as:

\begin{figure*}

\includegraphics[trim=1cm 4.0cm 0cm 5cm, clip=true, width=0.99\textwidth]{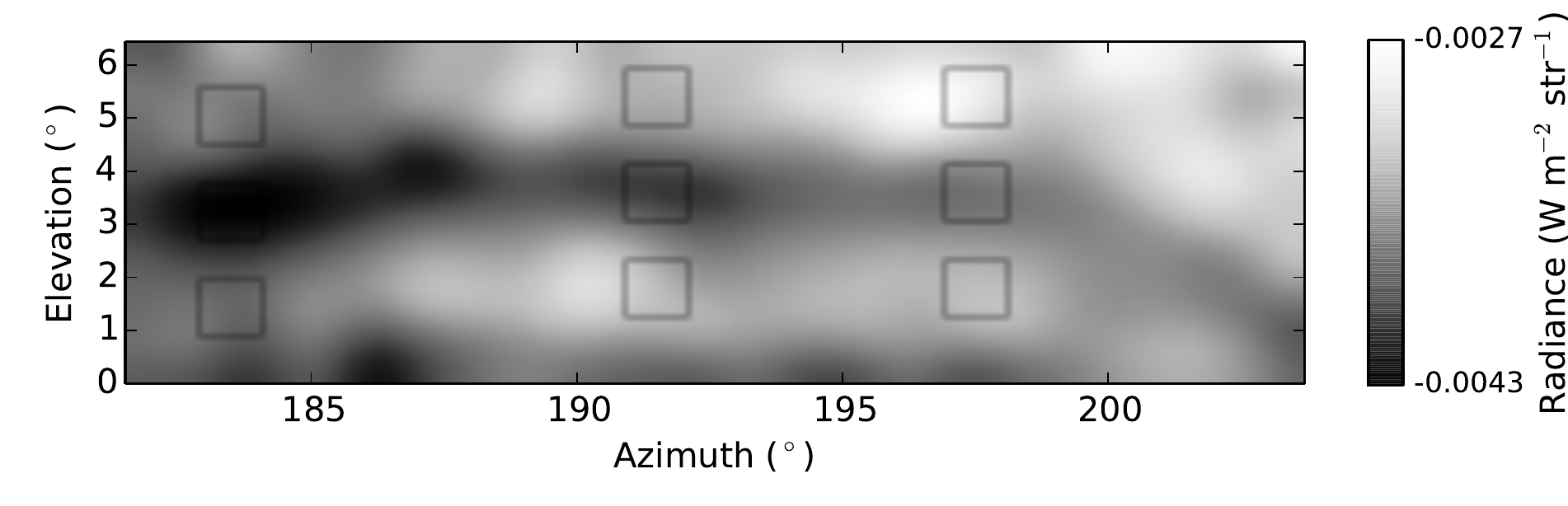}

\caption{Feature and background regions selected for optical depth measurements in image 0979.}\label{optcalcreg}
\end{figure*}

\begin{equation}
CL=\frac{1.96\sigma}{\sqrt{N}}
\end{equation}

\noindent where $\sigma$ is the standard deviation across the meanframe subtracted image and N is the number of pixels in each region (36). The standard deviation across the entire meanframe subtracted image was chosen rather than the standard deviation of the region as the former will take into account any larger scale variations in background across the frame. The confidence limits calculated in this fashion are in excellent agreement with estimated confidence limits from the signal-to-noise ratio of the DISR imager: the compression algorithm was designed to give an image with a mean signal-to-noise ratio of 100 (L. Doose, priv. comm.), thus from a 36 pixel region, where each pixel has radiance $\sim0.06$, the 95\% confidence limit on the mean value in the region is $\sim2\times10^{-4}$.

\subsection{Optical depth estimates}

Assuming that the feature is sufficiently optically thin that any change in radiance can be accounted for in a linear fashion, the change in optical depth of the feature with respect to its predicted value from the background regions, $\Delta\tau$, is given by:

\begin{equation}
\Delta\tau=-\ln\left(1-\frac{4\pi\Delta I}{F}\right)
\end{equation}

\noindent where $F$ is the Solar flux at the surface of Titan and $\Delta I$ is the difference between the feature and the predicted radiance from background regions in the mean-frame subtracted images. The calculated values of $\Delta I$ can be found in column 7 of Table \ref{RadTab}. The Solar flux was taken as 0.8123 W m$^{-2}$, as found from fitting the background radiance gradient - see Sect. \ref{rad_measurements}. 

The calculated values of $\Delta\tau$ are shown in the final column of Table \ref{RadTab}.

\section{Discussion}

\subsection{Reliability of detection}

\begin{figure*}
\addtolength{\subfigcapskip}{-3mm}
\centering

\subfigure[182.1$^{\circ}$ azimuth.]{\includegraphics[height=0.25\textheight]{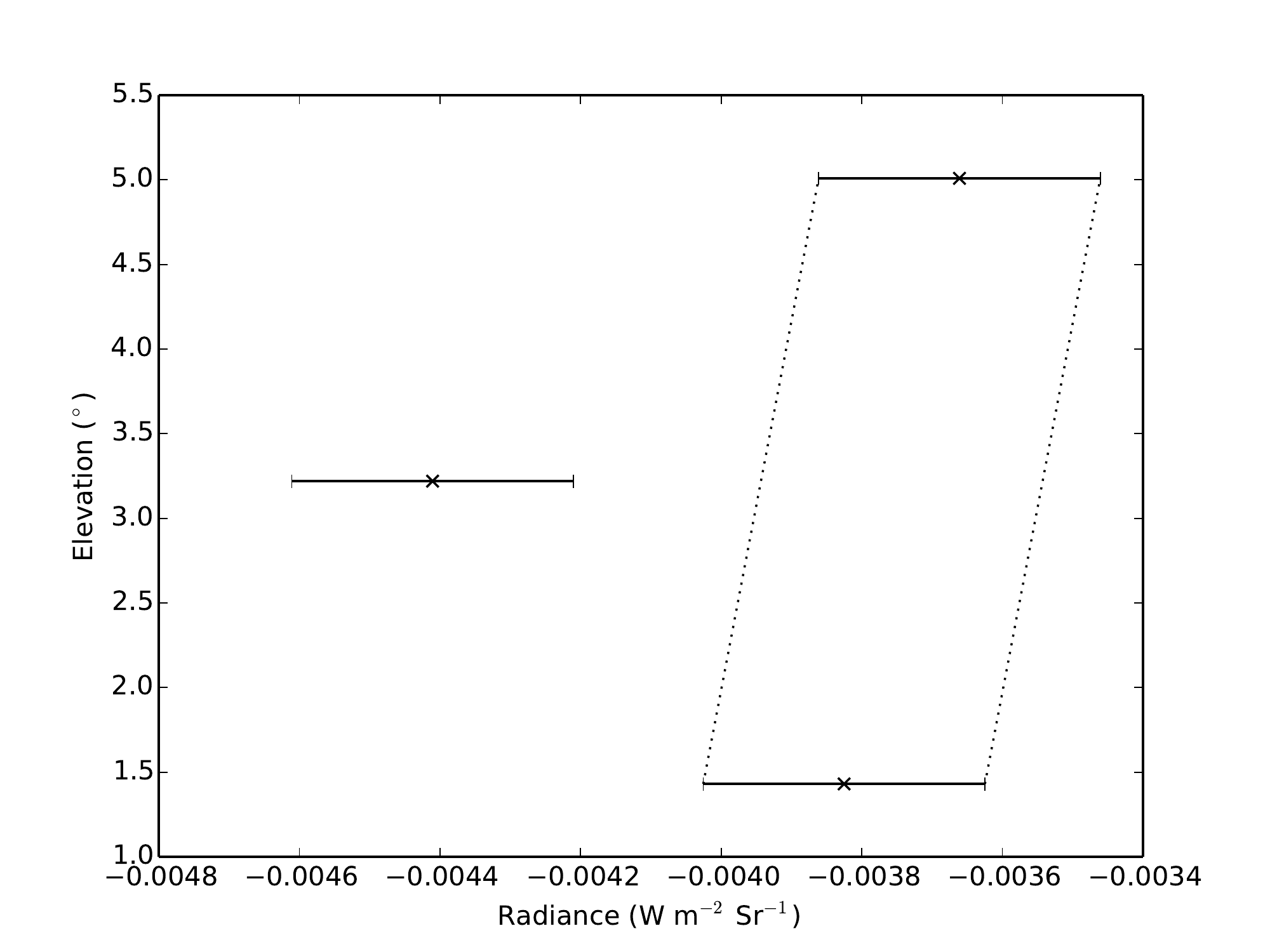}}

\subfigure[190.1$^{\circ}$ azimuth.]{\includegraphics[height=0.25\textheight]{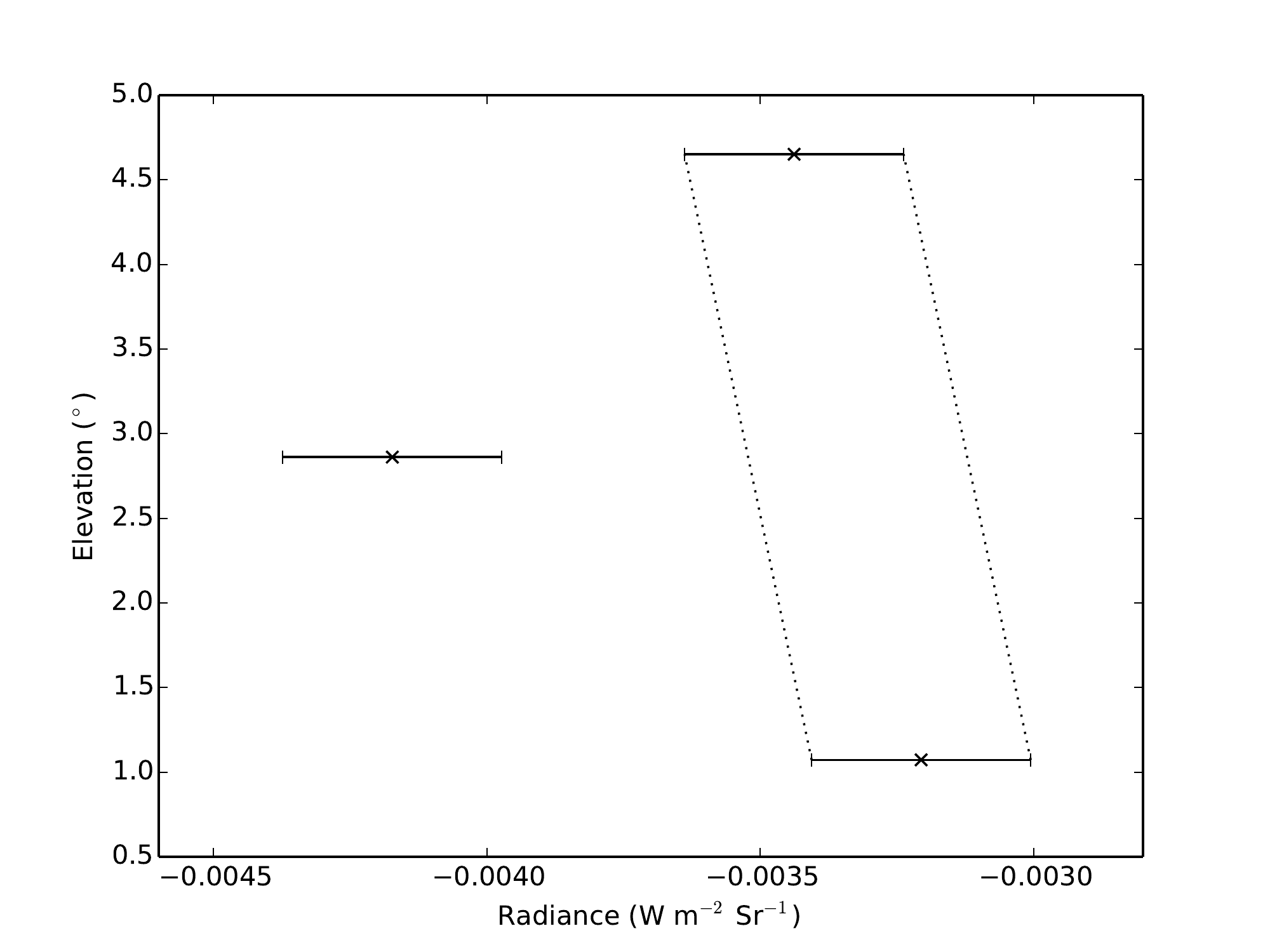}}

\subfigure[196.1$^\circ$ azimuth.]{\includegraphics[height=0.25\textheight]{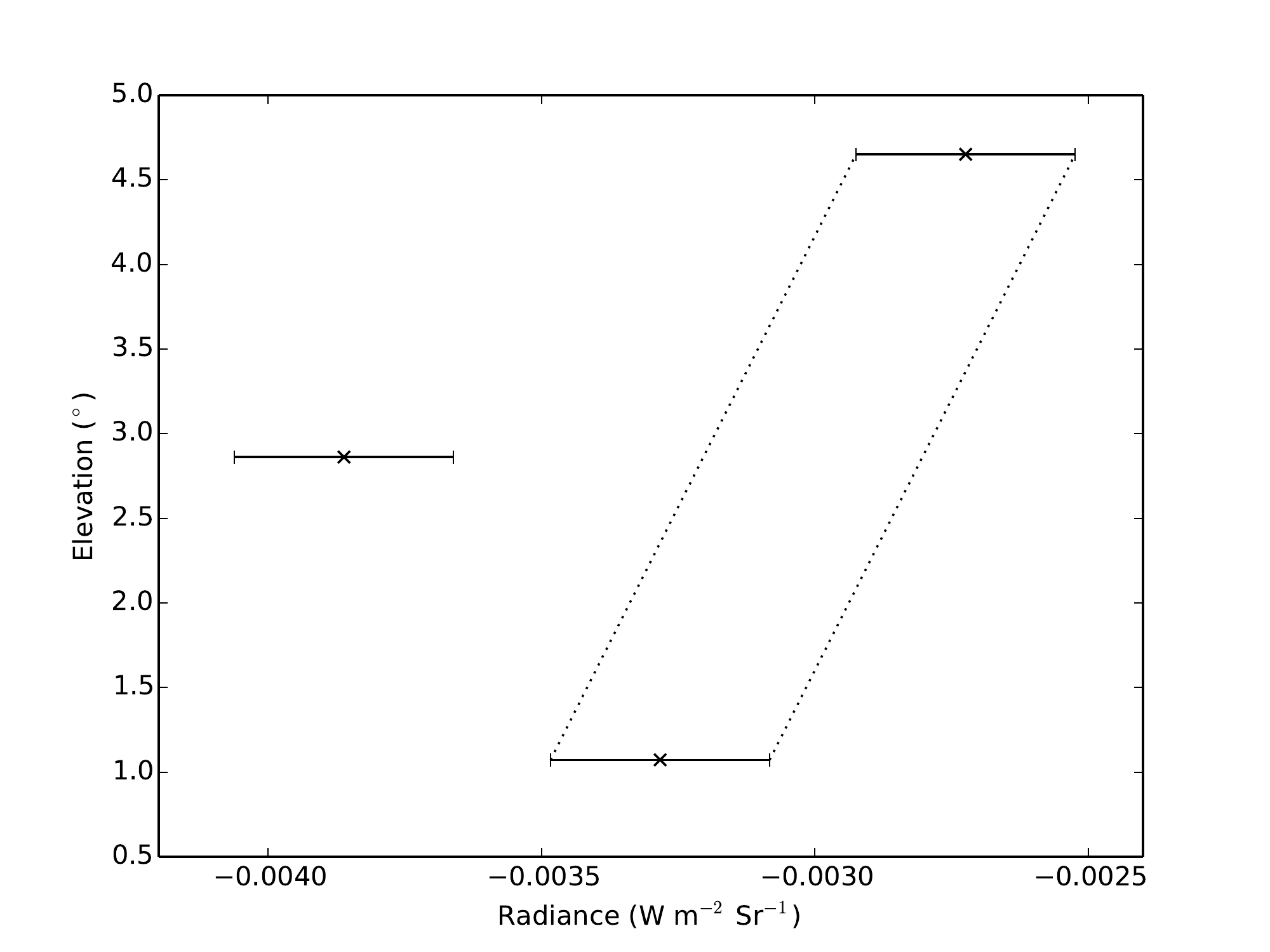}}

\caption{Radiance values of the feature (central elevation point in each image) compared to the predicted range from the background regions for image 0979 (MT 11185.34 sec). Images are all mean-frame subtracted with no smoothing}\label{fig_strength_plot}
\end{figure*}

In more than half of the selected feature regions with no mean frame subtraction or Gaussian smoothing, the radiance of the feature region lies more than the measured 95\% confidence interval from the predicted radiance from the mean background measurements. In addition, the feature is extended in every image, further compounding the reliability of detection. In the mean-frame subtracted images, in all cases the radiance of the feature lies outside the measured 95\% confidence interval of the predicted background level. These are shown for image 0979 (MT 11185.34 sec) in Fig. \ref{fig_strength_plot}. The regions referred to are shown in Fig. \ref{optcalcreg}. 

\begin{figure*}
\addtolength{\subfigcapskip}{-3mm}
\centering

\subfigure[Smoothed image.]{\includegraphics[width=0.8\textwidth]{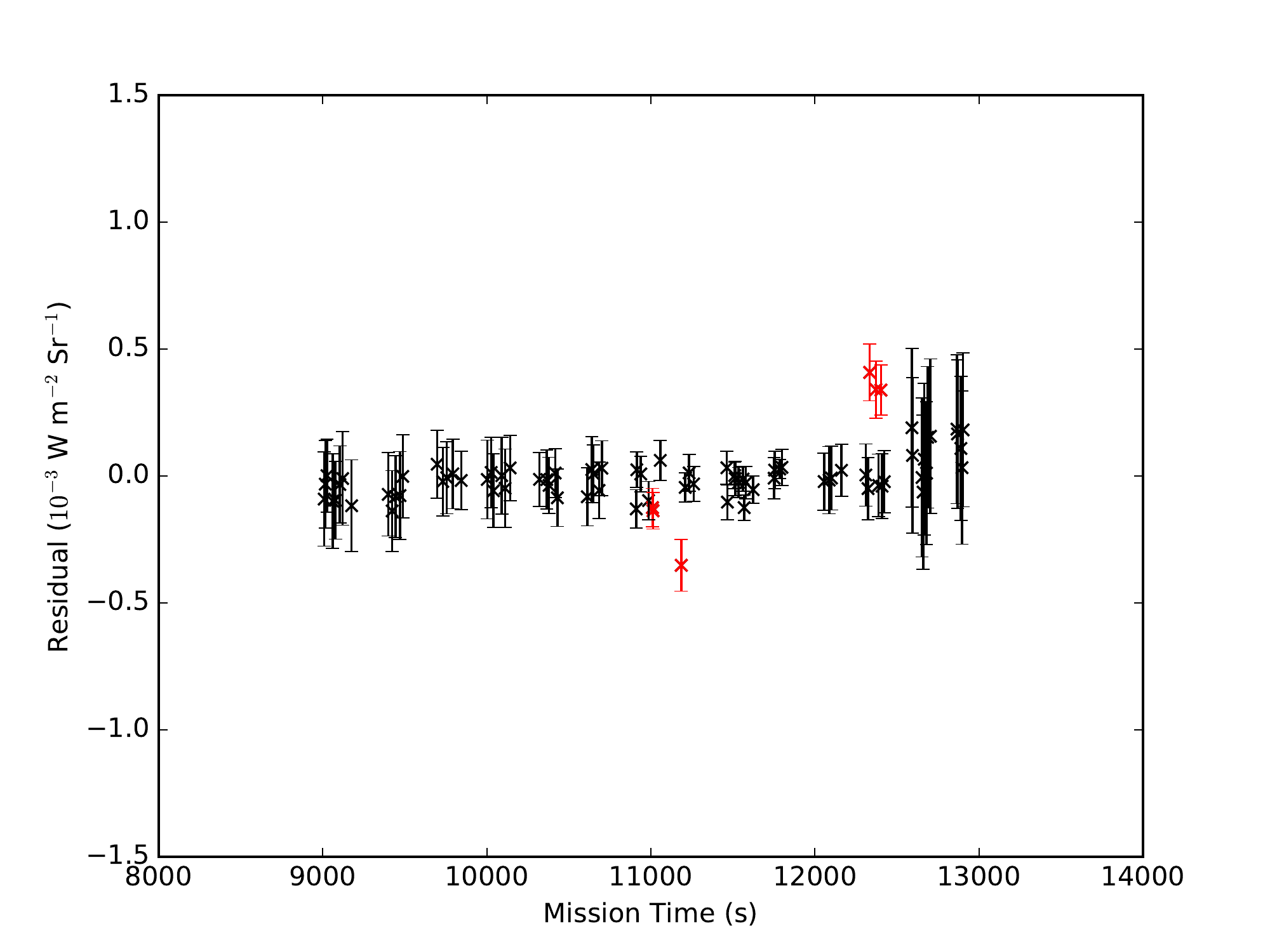}}
\subfigure[Unsmoothed image.]{\includegraphics[width=0.8\textwidth]{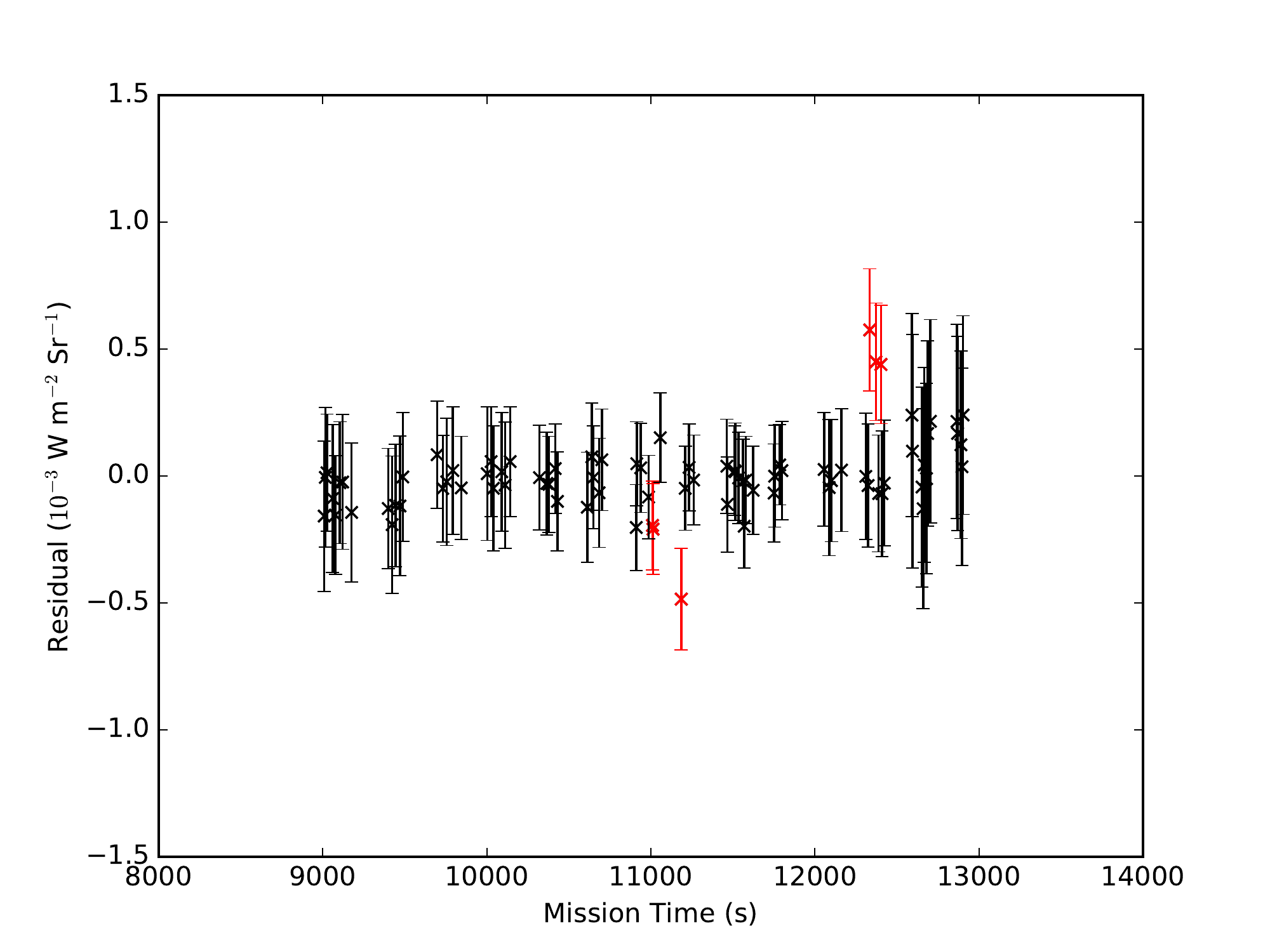}}

\caption{Plot of the feature radiance minus the predicted background radiance for all images since landing occurred, calculated from horizontal strips across the images (see text for further details). The red points indicate frames with detections. The initial two frames are not significantly different from zero due to their variation in the vertical direction.}\label{timeseries_plots}
\end{figure*}

\begin{figure*}
\addtolength{\subfigcapskip}{-3mm}
\centering

\subfigure[Control region above feature region, smoothed.]{\includegraphics[width=0.8\textwidth]{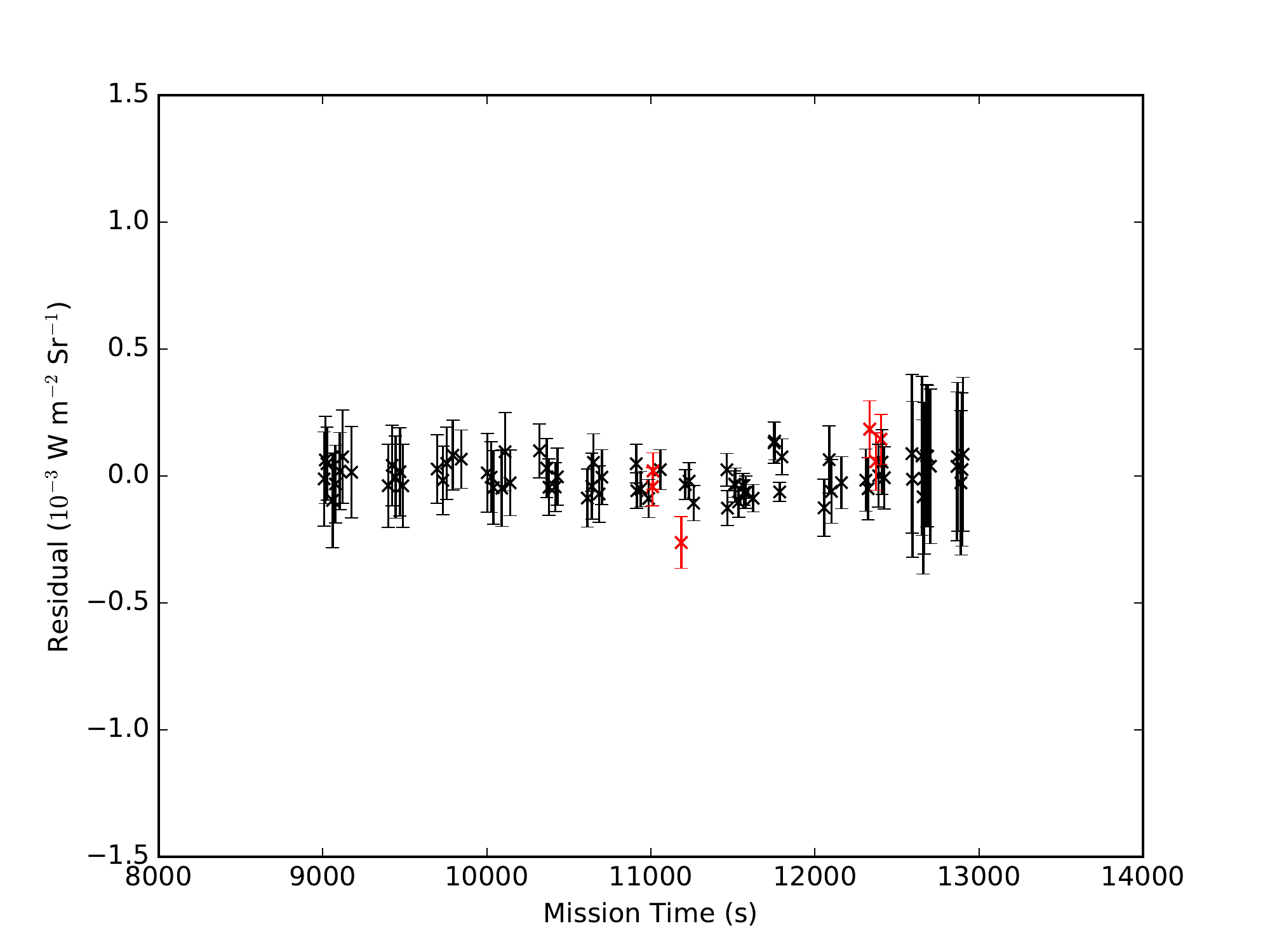}}
\subfigure[Control region below feature region, smoothed.]{\includegraphics[width=0.8\textwidth]{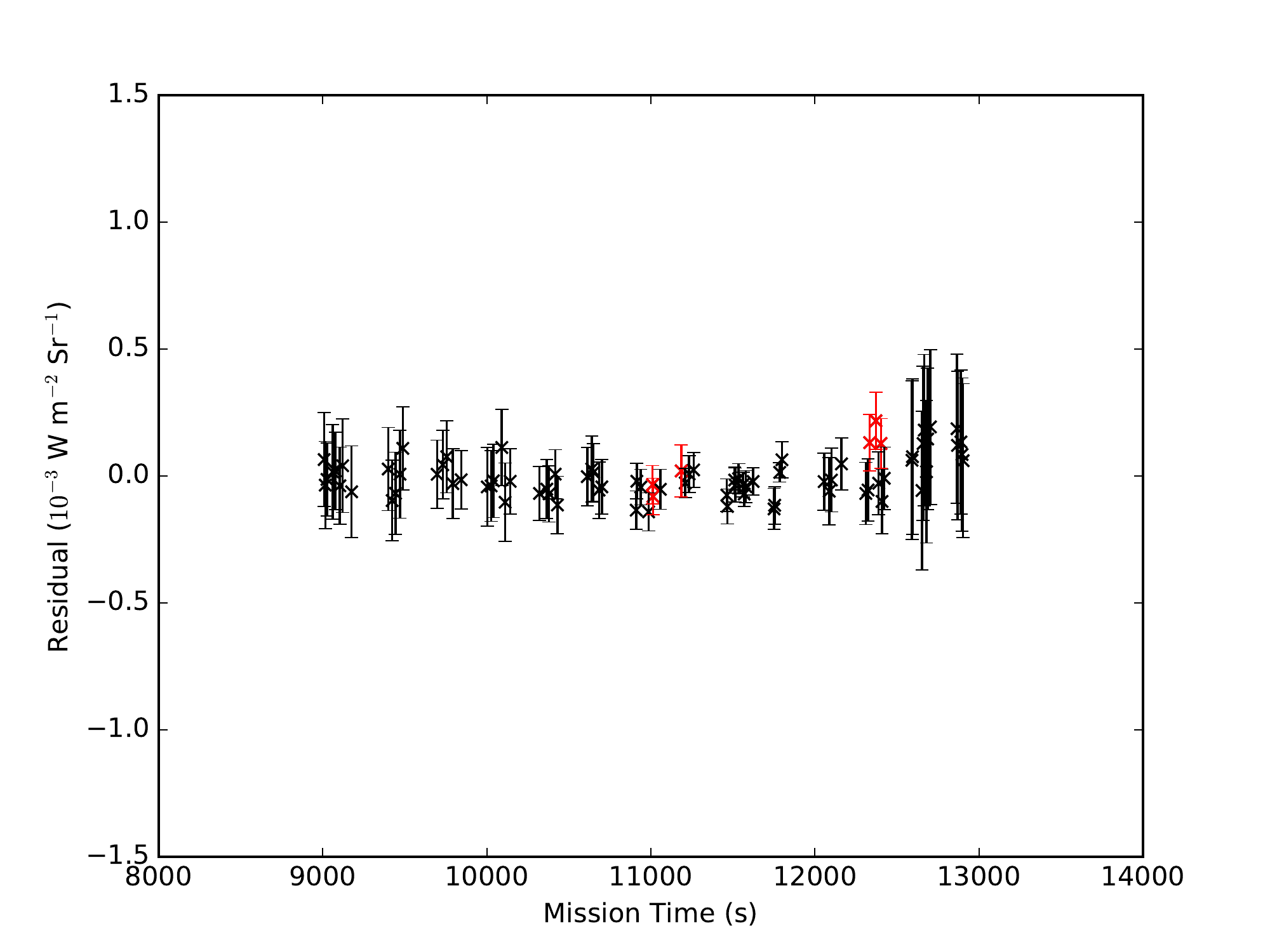}}

\caption{Plot of control region ``feature" radiance minus the predicted background radiance for all images since landing occurred, calculated from horizontal strips across the images (see text for further details). The control region is taken above the height of the feature region and the lower plot is taken below the height of the feature region. Background region offsets remain the same as for Fig. \ref{timeseries_plots}. A borderline detection in the upper plot is likely due to partial contamination of the control region with the feature. }\label{control_regions}
\end{figure*}

Comparisons of the detections as a function of time have also been investigated. Fig. \ref{timeseries_plots} shows the residual of the radiance of the feature region and the mean of two background regions centred 10 pixels above and below the feature region as a function of time. This time period covers the entire SLI ground imagery available. In this figure, for ease of visibility, the feature region has been taken as a horizontal strip across the entire image with a height of 6 pixels and its vertical position is held constant in all images. The background regions are also 6 pixel wide strips held constant across all images. The images with detections are shown in red with all other images shown in black. Four of the images have significantly non zero residuals averaged across the image strip (error bars state the 95\% confidence limits). The remaining two detections which are not significantly outside the zero residual region are the initial two detections (MT: 11009.76 and 11013.18 seconds) which have shapes that are more vertically variable, meaning that a single horizontal strip across the entire image does not capture primarily the feature region and rather a mixture of feature and non-feature regions. Time series plots of control regions are shown in Fig. \ref{control_regions}.

\begin{table*}
\centering

\caption{The absolute difference in radiance ($\Delta I$) between the measured feature (f) radiance and the mean of the two background (bg) regions at three azimuth angles per detection image are shown in the table below. The mean value of the three angles per image are also listed. The elevation of the central pixel of each region is indicated for both the feature and both background regions. The value in parentheses indicates the 95\% confidence limit on the final digit.\bigskip}\label{RadTab}
\renewcommand{\arraystretch}{0.9}

\begin{tabular}{c c c c c c}
\hline
Mission Time	&	Az.	&	El. (f)		& El. (bg)	& $\Delta I\times10^{4}$ 	&\multirow{2}{*}{$\Delta\tau$}\\
(s)		& ($^{\circ}$)	&	 ($^{\circ}$)	&  ($^{\circ}$)	&	(W m$^{-2}$ Sr$^{-1}$) 	&  \\
\hline
11009.76	&	182.5	&	3.4	&	5.2, 1.6	&	8(3)	&	0.012(5)\\
		&	188.9	&	3.8	&	5.5, 2.0	&	7(3)	&	0.011(5)\\
		&	197.5	&	3.2	&	5.0, 1.4	&	4(3)	&	0.006(5)\\
		&	\multicolumn{3}{c}{Mean}			&	6(2)	&	0.010(3)\\
11013.18	&	183.5	&	3.0	&	4.8, 1.3	&	3(3)	&	0.005(5)\\
		&	190.5	&	3.4	&	5.2, 1.6	&	6(3)	&	0.009(5)\\
		&	197.5	&	3.0	&	4.8, 1.3	&	5(3)	&	0.008(5)\\
		&	\multicolumn{3}{c}{Mean}			&	4(2)	&	0.007(3)\\
11185.34	&	183.5	&	3.2	&	5.0, 1.4	&	7(3)	&	0.010(5)\\
		&	191.5	&	3.6	&	5.4, 1.8	&	9(3)	&	0.013(5)\\
		&	197.5	&	3.6	&	5.4, 1.8	&	9(3)	&	0.013(5)\\
		&	\multicolumn{3}{c}{Mean}			&	8(2)	&	0.012(3)\\
12334.4		&	184.5	&	2.5	&	4.3, 0.7	&11(4)\phantom{1}&	0.016(6)\\
		&	192.5	&	3.6	&	5.4, 1.8	&	8(4)	&	0.013(6)\\
		&	197.5	&	2.9	&	4.7, 1.1	&	4(4)	&	0.007(6)\\
		&	\multicolumn{3}{c}{Mean}			&	7(2)	&	0.012(3)\\
12371.71	&	183.5	&	2.5	&	4.3, 0.7	&	9(4)	&	0.014(6)\\
		&	191.5	&	3.2	&	5.0, 1.4	&	5(4)	&	0.007(6)\\
		&	198.5	&	3.0	&	4.8, 1.3	&	5(4)	&	0.008(6)\\
		&	\multicolumn{3}{c}{Mean}			&	6(2)	&	0.010(3)\\
12402.39	&	183.5	&	2.9	&	4.7, 1.1	&	9(4)	&	0.013(6)\\
		&	191.5	&	3.0	&	4.8, 1.6	&	7(4)	&	0.010(6)\\
		&	198.5	&	2.9	&	4.7, 1.1	&	8(4)	&	0.012(6)\\
		&	\multicolumn{3}{c}{Mean}			&	8(2)	&	0.012(3)\\
\hline
\end{tabular}

\end{table*}

The feature could be an artifact of the detector or introduced during calibration, reduction or compression of the data. This is considered reasonably unlikely as detector features are extensively examined in \citet{CalibrationReport} and no features resembling this one are observed. Additionally, compression artifacts are seen in linear patterns as shown in Fig. \ref{mfsimage}, whereas this feature does not follow this pattern. An electrical wave across the CCD cannot be excluded as a potential cause of the feature, although this is fairly unlikely as it is only seen in a small subset of images.

\subsection{Other notable events after landing}

A number of other events of note have been reported both before and after the Huygens touchdown\footnote{A summary of these events, compiled by R. Lorenz, can be found at http://pds-atmospheres.nmsu.edu/data\_and\_services/atmospheres\_data/Huygens/extras/KeyEvents-3.htm}. The first feature detection is observed $\sim1000$ s after the potential dewdrop detection of \citet{Karkoschka2009} and at a similar time to the increase in CO$_2$ detected by GCMS \citep{Niemann2010}. The last feature detection is made $\sim$600 s prior to the end of the Cassini-probe link.

\subsection{Feature origin}

The observed feature has several potential origins. As clouds have been detected in Titan's atmosphere at a variety of altitudes, it is possible that this feature is caused by a low-level cloud. However, the feature has a similar morphology throughout the observations which are separated by up to 23 minutes and the feature does not appear to exhibit systematic movement across the field of view.  It has been shown (e.g. \citealp{Kloos2015} on Mars) that cloud movement can be detected from surface observations on time spans of less than a minute. The observations presented in this work are of lower resolution than \citet{Kloos2015}, but detectable, systematic movement would be expected. Taking a cloud height equal to the depth of the planetary boundary layer (300 m \citealp{Tokano2006}), a wind speed of 0.1 ms$^{-1}$ \citep{Tokano2006} and assuming the cloud moves perpendicular to the viewing direction, the time it would take a cloud to pass across the entire field of view is 1380 s. This is longer than the length of time between the two detection clusters. With a higher wind speed of 0.76 ms$^{-1}$, the meridional speed component of Huygens at 250m altitude reported by \citet{Karkoschka2015}, the time to traverse the frame entirely would be 182 s, which is shorter than the time between the two clusters of detections. Thus, this theory is unlikely but cannot be entirely discounted.

The observed feature could originate from a background hill or rise that is mostly obscured by the atmospheric haze. \citet{Karkoschka2007} has produced projections of the region surrounding the Huygens Landing (their Fig. 2). This figure shows a number of raised areas. The highlands in the north are estimated to be 150-200 m high, thus our feature, which would likely lie in the closer lower region, would have to be lower in elevation than this. The observed feature on the image lies approximately 25 pixels higher than the horizon. The whole image is 256 pixels in height and spans a field of view of 51\textdegree. From geometry, if the centre of the feature was at a radial distance from the probe of 50 m or 100 m and was caused by a vertical rise, it would have a vertical height of 4 m or 8 m above the observed ground respectively. Taking into account the 1-2 m rise at the horizon, the total height could be as high as 9 or 10 m above the ground at closest radial distance. The horizon, according to \citet{Karkoschka2007}, is  at a radial distance of 30 m from the probe, thus this would be a reasonable distance. A light feature is observed in Fig. 2 of \citet{Karkoschka2007} in the projected field of view of the DISR instrument at approximately this distance. Therefore this could be a reasonable explanation for the image feature, but atmospheric effects such as a change in fog opacity, would be required to hide and then display the feature.

The feature could be due to the presence of a superior mirage. Mirages form over surfaces that are either hotter or colder than the air above them, causing the density and subsequently the refractive index of the air to change with height. Above surfaces that are warmer than the air, inferior mirages can form: light rays are bent upwards towards the observer causing the image to appear below the object. Superior mirages, where the image appears above the object, are formed above surfaces that are colder than the surrounding air and thus require the presence of a temperature inversion. Further details on mirages can be found in \citet{Greenler1989} and \citet{Lehn92}. 

\begin{figure*}
\centering

\subfigure{\includegraphics[trim=0cm 0cm 3cm 17.75cm, clip=True, width=0.9\textwidth]{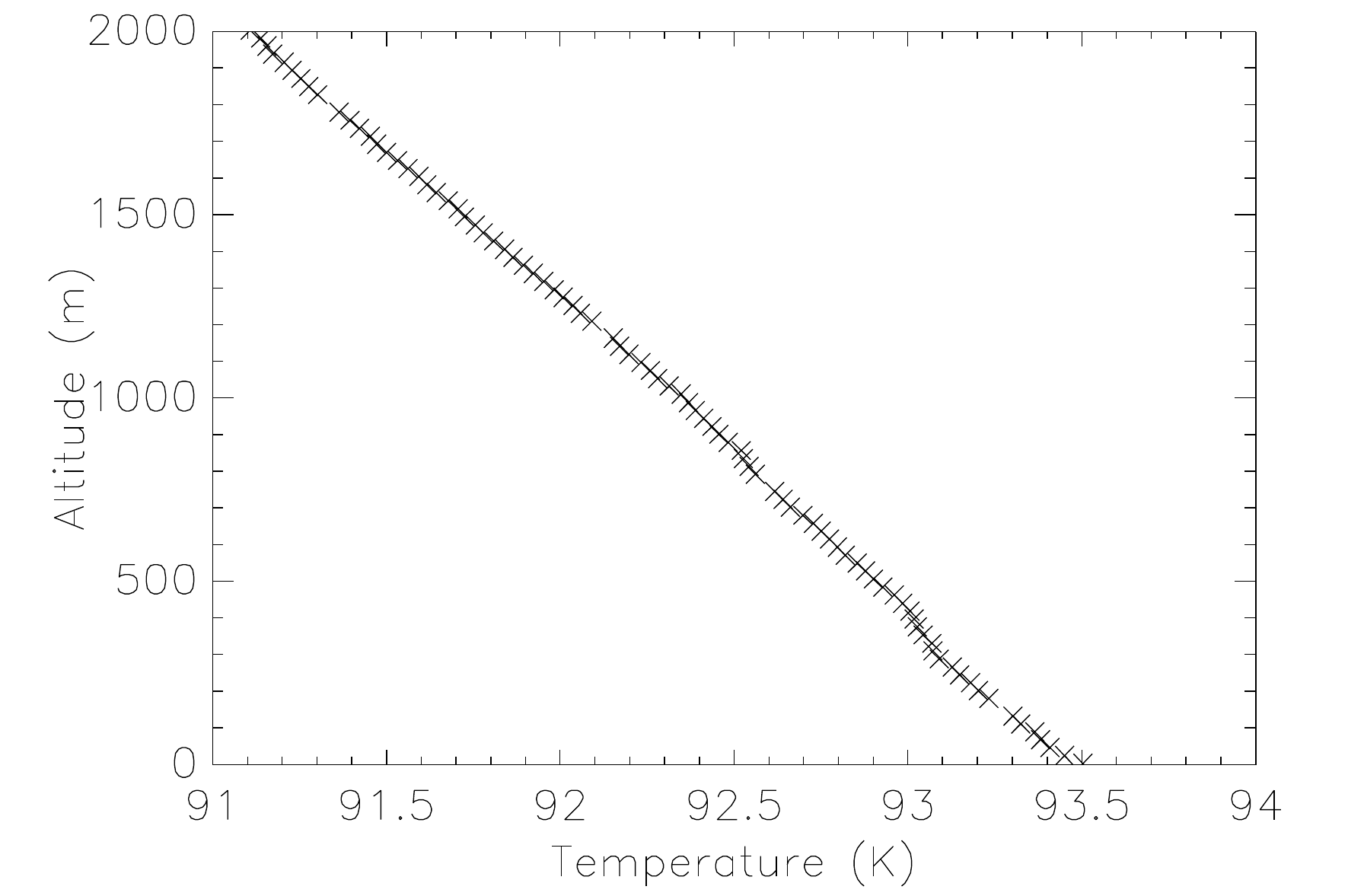}}

\bigskip

\bigskip

\subfigure{\includegraphics[trim=0cm 0cm 3cm 17.75cm, clip=True, width=0.9\textwidth]{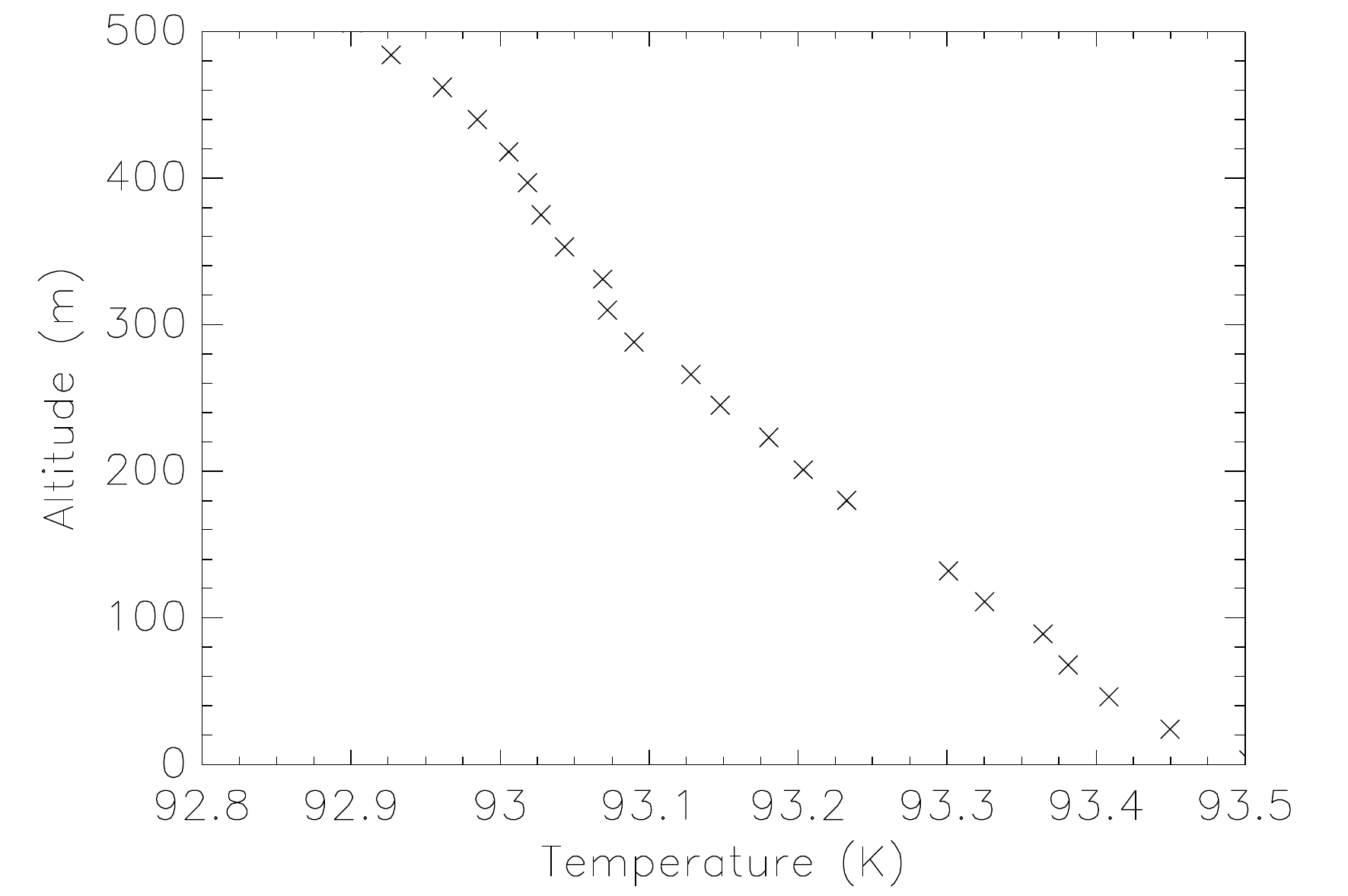}}

\caption{The change in temperature of the atmosphere of Titan as measured by the HASI instrument on-board Titan \citep{Fulchignoni2002} in the final 2km (upper plot) and final 500m (lower plot) of the descent of the Huygens probe.}\label{desc_temp}
\end{figure*}

The feature in the images of Titan's atmosphere, if a mirage, would be a superior mirage as it appears above the horizon. The Huygens probe measured the atmospheric temperature throughout the descent and would have detected a temperature inversion. The fully reduced temperature data, corrected for local atmospheric effects as described in \citet{Colombatti2004}, were taken from the European Space Agency's Planetary Science Archive\footnote{ftp://psa.esac.esa.int/pub/mirror/CASSINI-HUYGENS/HASI/HP-SSA-HASI-2-3-4-MISSION-V1.1/DATA/PROFILES/HASI\_L4\_ATMO\_PROFILE\_DESCEN.TAB}. The data for the last 2km and 500m of the descent are shown in Fig. \ref{desc_temp}.  As is shown, there is no detection of a temperature inversion and the surface temperature as measured by HASI is $93.65\pm0.25$ K \citep{Fulchignoni2005}, which is higher than the measured air temperature before the probe landed. It is possible, however, that a localised temperature inversion exists some distance from the probe's landing site which could cause the mirage. 

 Analyses of the Huygens' gas chromatograph mass spectrometer (GCMS) data by \citet{Lorenz2006} showed that the observed temperature evolution after impact on the surface of Titan of the heated GCMS inlet required a heat sink to be present. This is consistent with the heat sink produced if the ground in which the inlet of the GCMS was embedded was a sand or clay dampened by liquid hydrocarbons. The aforementioned potential dewdrop detection was made by \citet{Karkoschka2009} from surface imagery and the GCMS detected methane, ethane and acetylene, amongst other compounds, on the surface (see \citealp{Niemann2010} for further details). Evaporation of surface volatiles has been suggested as the cause of both an attenuation in signal strength observed by the speed-of-sound instrument \citep{Lorenz2014} and the sudden drop in apparent permittivity, detected with PWA-MIP \citep{Hamelin2015}. Evaporation of surface liquid could cause a loss of latent heat, leading to surface cooling similar to the situation over a lake where superior mirages are often seen on Earth. 

The most likely explanation for the observed feature is the presence of a fog bank. Fog had been hypothesised to exist on Titan and subsequently detected at its South Pole \citep{Brown2009}. A fog bank could vary reasonably quickly with time: from three-dimensional diffusion, the mean distance travelled by a particle is $\sqrt{6Dt}$ where $D$ is the diffusion coefficient ($7.4\times10^{-3}$  m$^2$s$^{-1}$, \citealp{Tokano2006}) and $t$ is the time in seconds \citep{Landau2013}. For $t=800$ s, the approximate time between the dark and light detections, the mean particle distance travelled is $\sim 6$ m or, at a distance of 30 m, a change in elevation angle of $11^\circ$. The calculated feature optical depths are also consistent with what would be expected from a fog bank or thin cloud \citep{Moores2015}. Mean-frame subtraction would remove the constant fog region leaving detections only at the upper edge when there is movement, likely resulting in a linear feature. The detections would cluster in time with the fog bank's movements: detections with the feature brighter than the background would appear when the fog bank falls and detections with the feature darker than the background would appear when the fog bank rises. This behaviour is observed in this work: the three images with dark features are seen clustered in time, the three images with bright features are seen clustered in time and there is a gap in detections of approximately 20 minutes between the two clusters.  The presence of the fog bank may persist for longer than the timespan of detections. In this case, the fog bank would be stationary in this time period and thus undetectable in mean frame subtraction.

The mean-frame would show the presence of a fog bank by a larger decrease in radiance with decreasing elevation compared with that expected from the atmospheric scattering. Fig. \ref{beersfit_average} shows that this is observed in the mean frame, with an average decrease in radiance of approximately 0.002-0.003 from the model value. This difference is similar to the change in radiance observed between the most intense feature regions and the predicted background, further supporting this theory.

\section{Conclusions}

This paper presents the results of an investigation into image data taken with the SLI imager on-board the Huygens probe after it landed on the surface of Titan in 2005. The data were calibrated according to the calibration report of \citet{CalibrationReport}, which involved correcting the dark current estimation carried out on-board the probe, correcting for detector sensitivity and dividing by the exposure time of the images.

Further processing was carried out over the entire image set including mean-frame subtraction, difference imaging and image co-adding to enhance any potential atmospheric features in the image. The images were blurred using a Gaussian filter to remove compression artifacts.

Out of the 82 images retrieved from the archive and taken with the Side Looking Imager, 6 contained an extended horizontal feature. This feature is only detectable in mean-frame subtracted or difference imaged frames. The intensity of the feature was measured in three $6\times6$ pixel regions and, for the majority of regions, the difference between the radiance of the feature compared with the predicted radiance from background regions was outside the 95\% confidence limit. The change in optical depths caused by the feature were in the range 0.005 to 0.014.

The presence of a fog bank rising and falling explains the change in feature intensity with respect to the background, the clustering of detections and the linear appearance of the feature in the mean-frame subtracted images. A fog bank also explains the difference between the predicted sky radiance in the non-mean frame subtracted images: the observed radiance of the sky decreases more than the predicted radiance. Therefore, for the aforementioned reasons the presence of a fog bank that rises and falls over the course of the observing period is considered the most likely explanation for the observed feature.

\section{Acknowledgements}

The authors would like to thank the European Space Agency and M. Tomasko, the Principal Investigator of the DISR instrument and M. Fulchignoni, the Principal Investigator of the HASI instrument for the use of data from the Huygens probe. This work was funded by the Natural Science and Engineering Research Council (NSERC) of Canada's Collaborative Research and Training Experience Program (CREATE) for Integrating  Atmospheric Chemistry and Physics from Earth to Space (IACPES). Finally, the authors would like to thank L. Doose for scientific discussions relating to this project and R. Lorenz and an anonymous reviewer for helpful comments and improvements for this paper. 

\bibliographystyle{elsarticle-harv}

\section{References} 
\bibliography{huygensbib}

\end{document}